\documentclass[aps,prc,preprint,groupedaddress,showpacs,floatfix]{revtex4}

\usepackage{epsfig}
\usepackage{amsmath}

\begin{document}

\preprint{OSU/Yanez-TKE-U235}

\title{Total kinetic energy release in the fast neutron-induced fission of $^{235}$U}

\author{R. Yanez, W. Loveland, J. King, J.S. Barrett}
\affiliation{Department of Chemistry, Oregon State University, Corvallis, OR 97331, USA}

\author{N. Fotiades, H.Y. Lee}
\affiliation{Los Alamos National Laboratory, Los Alamos, NM 87545, USA}

\date{\today}

\begin{abstract}

We have measured the total kinetic energy (TKE) release for the $^{235}$U(n,f) reaction for $E_{n}$=2-100 MeV using the 2E method with an array of Si PIN diode detectors. The neutron energies were determined by time of flight measurements using the white spectrum neutron beam at the LANSCE facility. To benchmark the TKE measurement, the TKE release for $^{235}$U(n$_{th}$,f) was also measured using a thermal neutron beam from the Oregon State University TRIGA reactor, giving pre-neutron emission $E^*_{TKE}=170.7\pm0.4$ MeV in good agreement with known values. Our measurements are thus absolute measurements. The TKE in $^{235}$U(n,f) decreases non-linearly from 169 MeV to 161 MeV for $E_{n}$=2-100 MeV. The multi-modal fission analysis of mass distributions and TKE indicates the origin of the TKE decrease with increasing neutron energy is a consequence of the fade out of asymmetric fission, which is associated with a higher TKE compared to symmetric fission. The average TKE associated with the superlong, standard I and standard II modes for a given mass is independent of neutron energy. The widths of the TKE distributions are constant from $E_{n}$=20-100 MeV and hence show no dependence with excitation energy.

\end{abstract}


\maketitle

\section{Introduction}
\label{intro}

The total kinetic energy release (TKE) in the neutron induced fission of $^{235,238}$U and $^{239}$Pu decreases with increasing incoming neutron energy \cite{akimov71,meadows82,muller84,straede87,zoller95,yanez14,meierbachtol16,duke16}. The rate of decrease is of the order of a few hundred keV per MeV incident neutron energy and has been tentatively attributed to the steady growth of symmetric fission \cite{madland06}, although some assert that the rate of change from asymmetric to symmetric fission may be too slow for it to be the cause \cite{lestone14}. The main contribution to the TKE is the Coulomb repulsion between the deformed fragments at scission. The nascent fragments may also reach the scission point with a kinetic energy, acquired during the descent from the saddle to the scission point, that has to be added to the total kinetic energy. The origin of the decrease of TKE with increasing neutron energy could be the result of changes in either, or both, or other effects. The fraction of symmetric fission increases as shell structure effects are washed out with increasing excitation energy, resulting in a lower overall TKE (symmetric fission is known to have a lower associated TKE release \cite{unik64}.) On the other hand, the increase of nuclear friction with excitation energy may be responsible for a decrease of the pre-scission kinetic energy, being dissipated into internal energy during the descent from the saddle to the scission point, also resulting in a lower overall TKE \cite{brosa90}. The physical phenomena responsible for the width (standard deviation) of the TKE distribution are poorly studied. It is believed to be temperature dependent, in which temperature driven fluctuations of the dissipation of pre-scission kinetic energy and the pre-scission inter-fragment separation are direct causes \cite{wilkins76,grossmann88}. In general, theories of fission have seldom tackled the issue of variances of TKE distributions. It has been recognized since the very first measurements of TKE that the experimental conditions of the measurement artificially increase the width of the distribution. Removing this contribution presents a challenge to experimenters. Hence, finding systematic trends that can experimentally relate the variance to fluctuations inherent to the fission process has not been straightforward. In this article, we report the first complete measurement of the total kinetic energy release in the neutron induced fission of $^{235}$U for the neutron energy range of $E_{n}$=2-100 MeV, a subject of great technological and scientific interest. Because of our large, ``high statistics" data set, we are able to examine questions of the widths of the TKE distributions, and the relative yields of symmetric and asymmetric fission as a function of neutron energy. 

This article is organized as follows: In Section~\ref{expt} we describe the experimental details. In Section~\ref{analysis}, the analysis method is described. In Section~\ref{results} we present the results of the measurements, which we discuss in Section~\ref{discussion}. The conclusions are presented in Section~\ref{conclusions}. In support of the many details of the analysis method we developed a Monte Carlo detector response simulation, which is described in Appendix~\ref{appA}.

\section{The experiment}
\label{expt}

The experiment was carried out at the Weapons Neutron Research Facility (WNR) at the Los Alamos Neutron Science Center (LANSCE) at the Los Alamos National Laboratory (LANL). White spectrum neutron beams were generated from an unmoderated tungsten spallation source using the 800 MeV proton beam from the LANSCE linear accelerator. The experiment was located on the 15R beam line (15$^\circ$ to the right with respect to the proton beam). The proton beam is pulsed allowing one to measure the time of flight of the neutrons arriving at the experimental area. The proton beam intensity was typically 1.8 $\mu$A. A fission ionization chamber \cite{wender93} located at the exit of the 1 cm diameter collimator was used to continuously monitor the absolute neutron beam intensities. At the entrance of the scattering chamber, the beam diameter was measured to be 1.0 cm (FWHM) with a photographic emulsion plate. 

The $^{235}$U target and the fission detectors were housed in an evacuated aluminum scattering chamber. The scattering chamber was located 55 cm from the collimator and $\sim 14$ m from the neutron beam dump. The center of the scattering chamber was located 13.85 m from the production target.

The $^{235}$U target consisted of a deposit of $^{235}$UF$_4$ on a C backing. The thickness of the $^{235}$U was 175.5 $\mu$g $^{235}$U/cm$^2$ while the backing thickness was 100 $\mu$g/cm$^2$. The isotopic purity of the $^{235}$U was 98.12\%. The target was tilted at 45$^\circ$ with respect to the incident beam.

\begin{figure}[hbp]
\includegraphics[trim=0cm 10cm 5cm 5cm,scale=0.6]{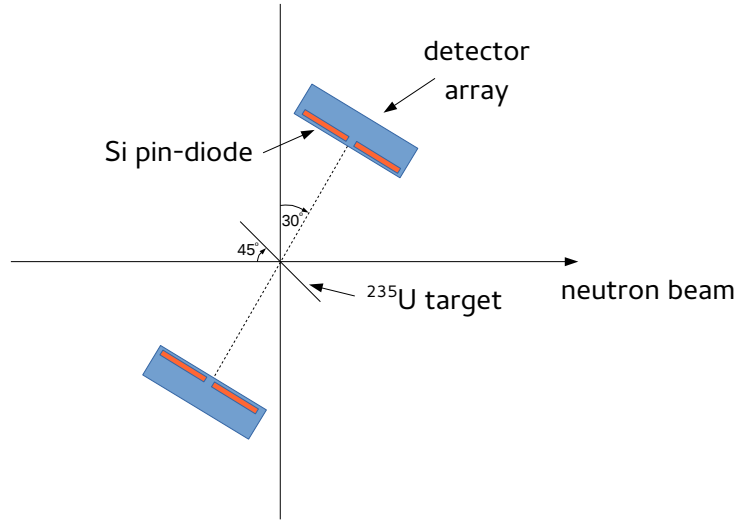}
\caption{Schematic illustration of the detector arrangement.}
\label{fig_setup}
\end{figure}

Fission fragments were detected in two arrays, on opposite sides of the beam, each consisting of four Si PIN photo-diodes (Hamamatsu S3590-09), arranged in a 2$\times$2 configuration, as close to each other as physically possible. The area of the individual PIN diodes was 1 cm$^2$. The distance from the center of the target to the center of the detectors was 2.1 cm. Fig.~\ref{fig_setup} depicts the detector arrangements. The detector arrays were positioned 60$^\circ$ and 120$^\circ$ with respect to the neutron beam. The arrangement is such that there are four pairs of detectors with an angle of 180$^\circ$ with respect to each other. The energy calibration of the fission detectors was done with a $^{252}$Cf source, which had a 50 $\mu$g/cm$^2$ Au cover.

The time of flight of each interacting neutron was measured using the timing pulse from a Si PIN diode and the accelerator RF signal. Absolute calibrations of this time scale were obtained from the photo-fission peak in the fission time spectra and the known flight path geometry. The error of the incident neutron energy was estimated with the width of the photo-fission peak, which was $\sim 1$ ns (standard deviation). The timing resolution of the detectors is 200 ps (Appendix~\ref{appA}.)

To benchmark the experimental method, the TKE in the $^{235}$U(n$_{th}$,f) reaction was measured at the Oregon State University 1 MW TRIGA Reactor. The measurement was made with the same apparatus and target as in the LANSCE experiment. The beam size was $\sim 4$ cm in diameter and the distance from the center of the target to the detectors were increased to 4.3 cm. Energy calibrations were done with a window-less $^{252}$Cf source. The thermal neutron flux incident on the target was $\sim 2.8 \times 10^7$ cm$^{-2}$s$^{-1}$ at 1 MW.

\section{Analysis}
\label{analysis}

\begin{figure}[hbp]
\includegraphics[scale=0.5]{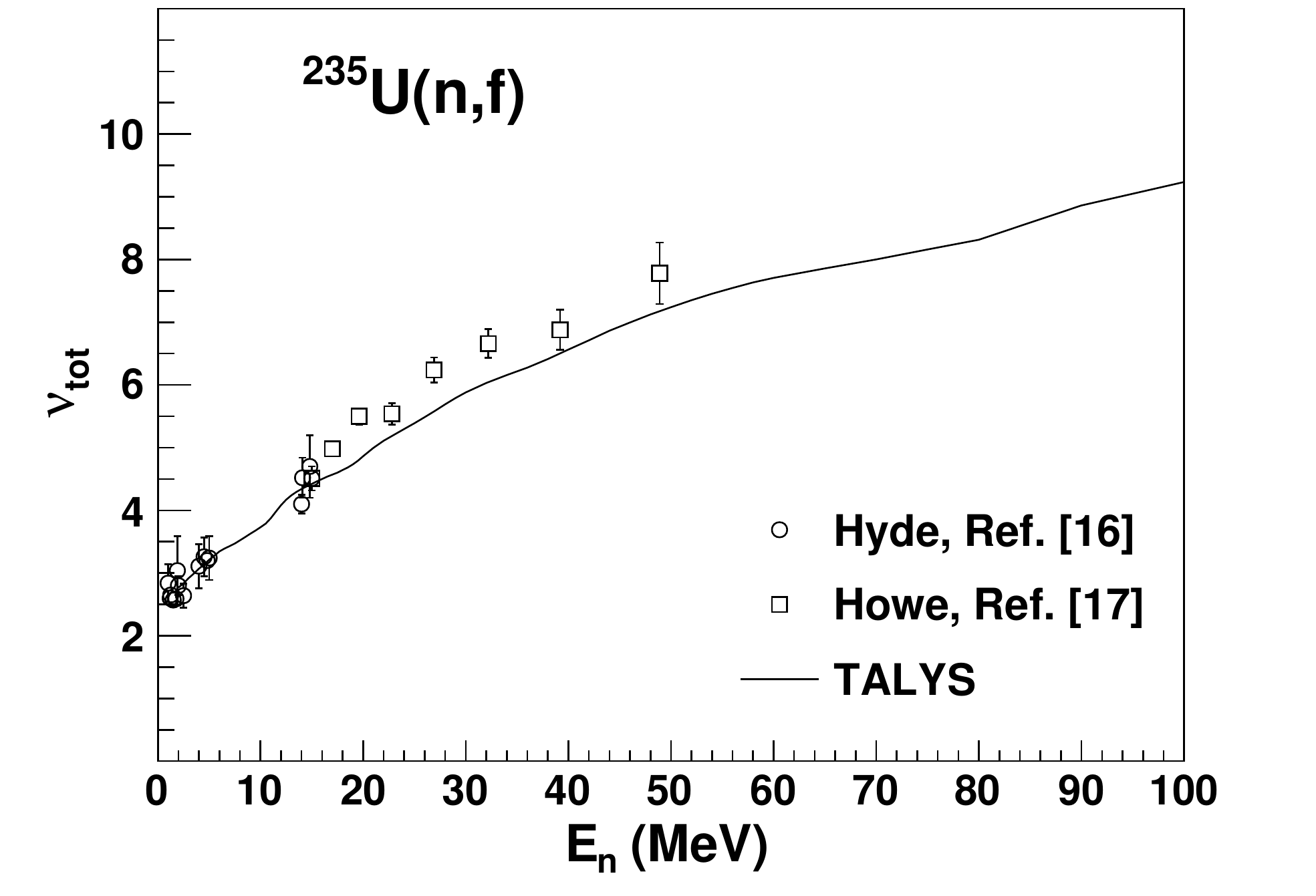}
\caption{Prompt neutron multiplicities calculated with TALYS \cite{talys}. Experimental data from Hyde \cite{hyde64} and Howe \cite{howe84} are plotted for comparison.}
\label{fig_nutot}
\end{figure}

The analysis of the data is based on the $2E$ method, which derives from the laws of mass and momentum conservation,
\begin{eqnarray}
\label{eq_kin}
A^*=A_A^*+A_B^*\\
\label{eq_kin2}
M_A^*E_A^*=M_B^*E_B^*
\end{eqnarray}
\noindent
where $A$ is the mass number of the fissioning nucleus, $A_{A,B}$, $M_{A,B}$ and $E_{A,B}$ are the fragment mass number, mass and kinetic energy, respectively. Quantities labeled with an asterisk refer to pre-fission quantities to clearly distinguish them from the post-neutron emission quantities. In the first step of the iterative analysis procedure, the mass number of the post-neutron emission fragments are assumed to be,
\begin{equation}
A_A=A_B=\frac{236-\nu_{tot}(E_n)}{2}
\end{equation}
\noindent
where $\nu_{tot}(E_n)$ is the total prompt neutron multiplicity. $\nu_{tot}(E_n)$ was estimated by using the TALYS \cite{talys} code, which defaults to the GEF code \cite{gef} when calculating the prompt-neutron multiplicities. In Fig.~\ref{fig_nutot} we show the TALYS calculation and the experimental data compiled by Hyde \cite{hyde64} and the data of Howe \cite{howe84}. The experimental data is well described by the calculation. The post-neutron emission fission fragment kinetic energies are calculated with the Schmitt procedure \cite{schmitt65,weissenberger} and corrected for energy losses in materials using the range correlations of Northcliffe and Schilling \cite{northcliffe}). The atomic number of the fission fragments are estimated by deducing the most probable atomic number $Z_{mp}$ from the measured independent yields in the $^{235}$U(n,f) reaction for 14.7 MeV neutrons \cite{endf-349}. We deduce the most probable charge by making Gaussian fits to $\sigma(A)$ and then use a linear fit to deduce $Z_{mp}(A)$. For the calibrations, the independent yields of $^{252}$Cf(SF) gives,
\begin{equation}
Z_{mp}=1.572+0.3812 A
\end{equation}
\noindent
The slope of the fit is rather close to $Z/A$ of $^{252}$Cf, but the offset is non-zero. It was determined with the help of the Monte Carlo detector response simulation (Appendix~\ref{appA}) that using $Z_{mp}$ over the assumption that the fragments preserve the $N/Z$ of the fissioning nucleus improved the mass resolution $\Delta m$ by 0.2 u. Similarly, the $Z_{mp}$ of the $^{235}$U(n,f) reaction for 14.7 MeV neutrons is,
\begin{equation}
Z_{mp}=1.503+0.3831 A
\end{equation}
\noindent
Having a slight advantage, we use $Z_{mp}(A)$ to deduce the fragment $Z$ needed to correct for energy losses, assuming $Z_{mp}$ in the $^{235}$U(n,f) reaction at 14.7 MeV is representative of fast neutron induced reactions with $^{235}$U.

\begin{figure}[hbp]
\includegraphics[scale=0.5]{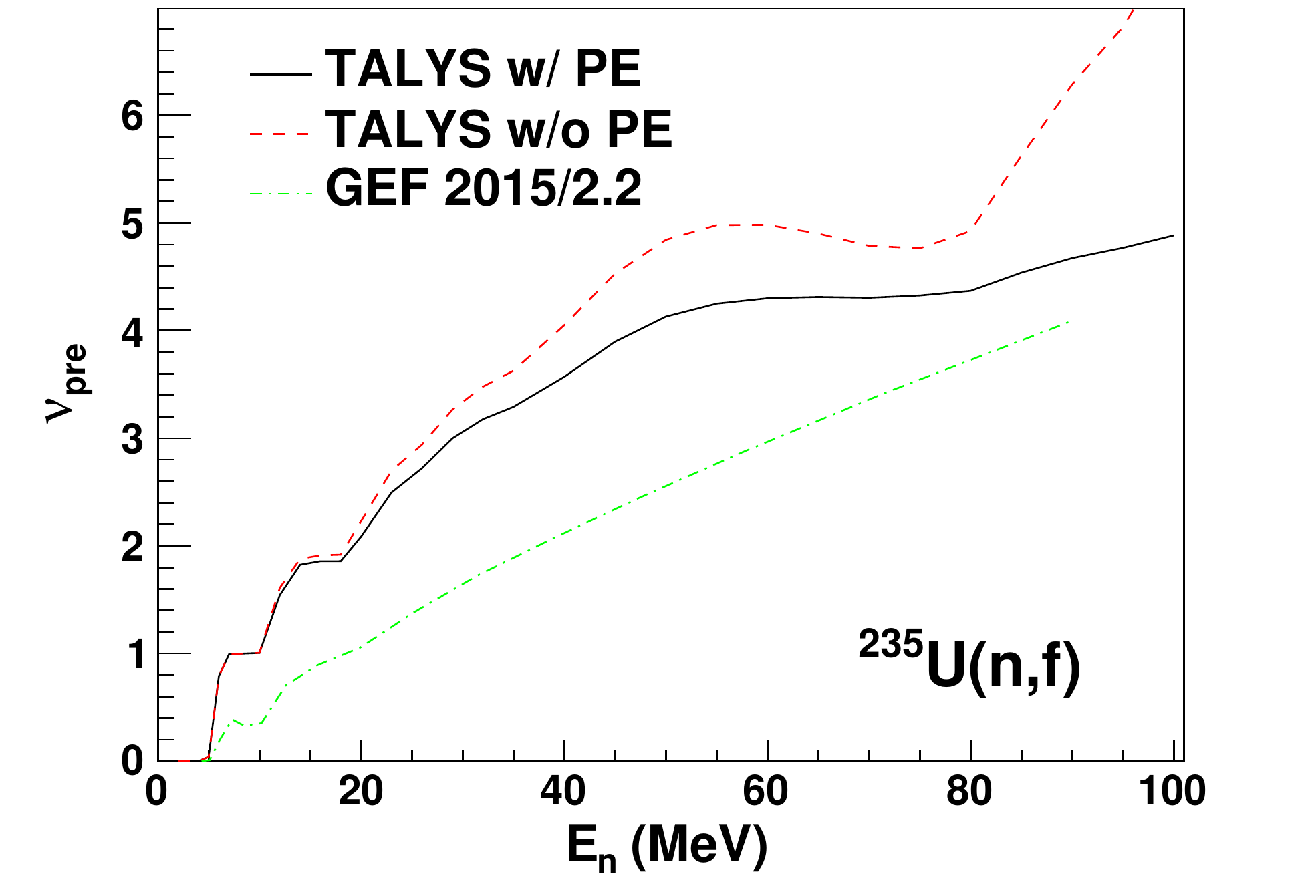}
\caption{Pre-fission neutron multiplicities calculated with TALYS \cite{talys} including pre-equilibrium emission (solid line), excluding pre-equilibrium emission (dashed line) and with the GEF code \cite{gef} (dot-dashed line).}
\label{fig_nupre}
\end{figure}

The pre-fission masses, $A_A^*$ and $A_B^*$, of a coincident pair of fission fragments are calculated by assuming isotropic neutron emission from the fully accelerated fragments in their respective center-of-mass (c.m.) frames. If emission is isotropic, the fragment velocities are, on average, unaffected by the recoils,
\begin{equation}
v_{A,B}^*=v_{A,B}
\end{equation}
\noindent
Hence,
\begin{equation}
A_{A,B}^*= \frac {v_{B,A}} {v_{A}+v_{B}} A^*
\end{equation}
\noindent
$v_{A,B}$ are calculated in the c.m.\ frame to account for the small momentum transfer given to the compound nucleus by the incoming neutron. $A_A$ and $A_B$ are iteratively varied, constrained by mass conservation,
\begin{equation}
A_A+A_B=236-\nu_{tot}(E_n)
\end{equation}
\noindent
to conserve momentum in the c.m.\ frame in the pre-fission stage. $A^*=236-\nu_{pre}(E_n)$, where $\nu_{pre}(E_n)$ is the pre-fission neutron multiplicity. This quantity has not been measured. We estimate $\nu_{pre}(E_n)$ by using the TALYS \cite{talys} code. In these calculations, we consider pre-equilibrium emission as a possible reaction mechanism. In Fig.~\ref{fig_nupre} we show the calculated $\nu_{pre}(E_n)$. The iterations are considered to have numerically converged when,
\begin{equation}
M_A^*E_A^*-M_B^*E_B^*=0
\end{equation}
\noindent
to better than one part in $10^5$. If the iterative procedure is done correctly, the average pre-fission mass distribution in a coincident pair of detectors must be equal, $\left<A_A^*\right>=\left<A_B^*\right>$, which is used as a consistency check. Mass numbers are treated as real numbers in the numerical solution of Eq.~\ref{eq_kin} and Eq.~\ref{eq_kin2}, but are truncated to the nearest integer number after convergence. The time difference between two coincident fission fragments was used to reject random coincidences. A time difference window of $|t_A-t_B|<10$ ns was imposed. About $\sim 0.5$\% of events were rejected by this condition.

With the aid of Monte Carlo detector response simulations (Appendix~\ref{appA}) it was determined that the average value of TKE is unaffected by the analysis procedure and detector geometry when all the corrections to the data are applied in the correct order. However, the geometry and analysis procedure do increase the width of the TKE distribution by $\sigma_{inst}=3.24$ MeV (instrumental standard deviation). The measured standard deviation $\sigma_{expt}$ is taken to be given by,
\begin{equation}
\sigma_{expt}^2 = \sigma_{TKE}^2 + \sigma_{inst}^2
\end{equation}
\noindent
where $\sigma_{TKE}$ is the standard deviation of the distribution independent of experimental conditions. Here we assume the cross terms in the covariance matrix are negligible; the variables upon which $\sigma_{TKE}$ and $\sigma_{inst}$ depend are not interconnected. The Monte Carlo simulations reveal basic features about $\sigma_{inst}$. For example, the aperture of the cone of emission, which intends to mimic recoil effects due to evaporation from the fission fragments, changes $\sigma_{TKE}$ but leaves $\sigma_{inst}$ unchanged, as it should. The energy resolution of detectors has no effect on $\sigma_{expt}$, whereas the most significant contribution to $\sigma_{inst}$ stems from the $dE/dx$ corrections made with mass numbers with resolution $\Delta m$ and change $Z_{mp}$. The pure geometrical effects are such that increasing the detector distance decreases $\sigma_{inst}$, whereas increasing the beam width increases $\sigma_{inst}$.

As mentioned in the Introduction (Section~\ref{intro}), the artificial increase of the measured TKE widths and its correction presents a real challenge to experimenters. There is a necessity to apply a reproducible correction method that is applicable to all measurements. To this end, we have measured the width of the TKE distribution of $^{252}$Cf(SF) using a thin source and the detector setup used in the $^{235}$U(n,f) experiment. The thin $^{252}$Cf source was made by evaporating 5 $\mu$l of a 25 nCi/$\mu$l solution on a 100 $\mu$g/cm$^2$ C foil, and had an area of 5.4 mm$^2$. The measured $\sigma_{expt}$ for $^{252}$Cf, integrated over TKE and fragment mass is 12.54 MeV. The same measurement was made using the slightly different detector setup used in the $^{235}$U(n$_{th}$,f) experiment. In this case the $\sigma_{expt}$ was measured to be 12.66 MeV. As reference we use the standard deviation of $^{252}$Cf measured by Schmitt \textit{et al.} \cite{schmitt66}, $\sigma_{TKE}=12.0$ MeV. With this we conclude that $\sigma_{inst}=3.65$ MeV in the $^{235}$U(n,f) experiment, and $\sigma_{inst}=4.04$ MeV in the $^{235}$U(n$_{th}$,f) experiment. The advantage of using the measured value of $\sigma_{TKE}$ of $^{252}$Cf as a reference to correct for $\sigma_{inst}$ is that the reported experimental standard deviations in the present TKE measurement can be adjusted at any time to another reference value of choice, and hence made comparable to other measurements.

The uncertainty of the TKE distribution mean and width has been estimated by varying the thicknesses of each energy degrading material by 5\%, the detector and target angles by $2^{\circ}$, the detector distance by 1 mm, and $\nu_{tot}$ and $\nu_{pre}$ by 5\%. This uncertainly, which intends to account for systematical errors, is added in quadrature to the statistical error.

The yield pattern of coincident fission fragments in the experimental data (there are 16 possible combinations) as compared to Monte Carlo detector response simulations is consistent with a displacement of the intensity of the beam by 2 mm horizontally from the center of the target. Although the experimental assembly was optically aligned with a laser beam, it was later found that the markers used to align the laser relative the beam line were misplaced, effectively displacing the beam from the center of the target by 2 mm as revealed by the simulations. The analysis of the data took into consideration this experimental condition.

\section{Results}
\label{results}

\subsection{The total kinetic energy}

\begin{figure}[hbp]
\includegraphics[scale=0.5]{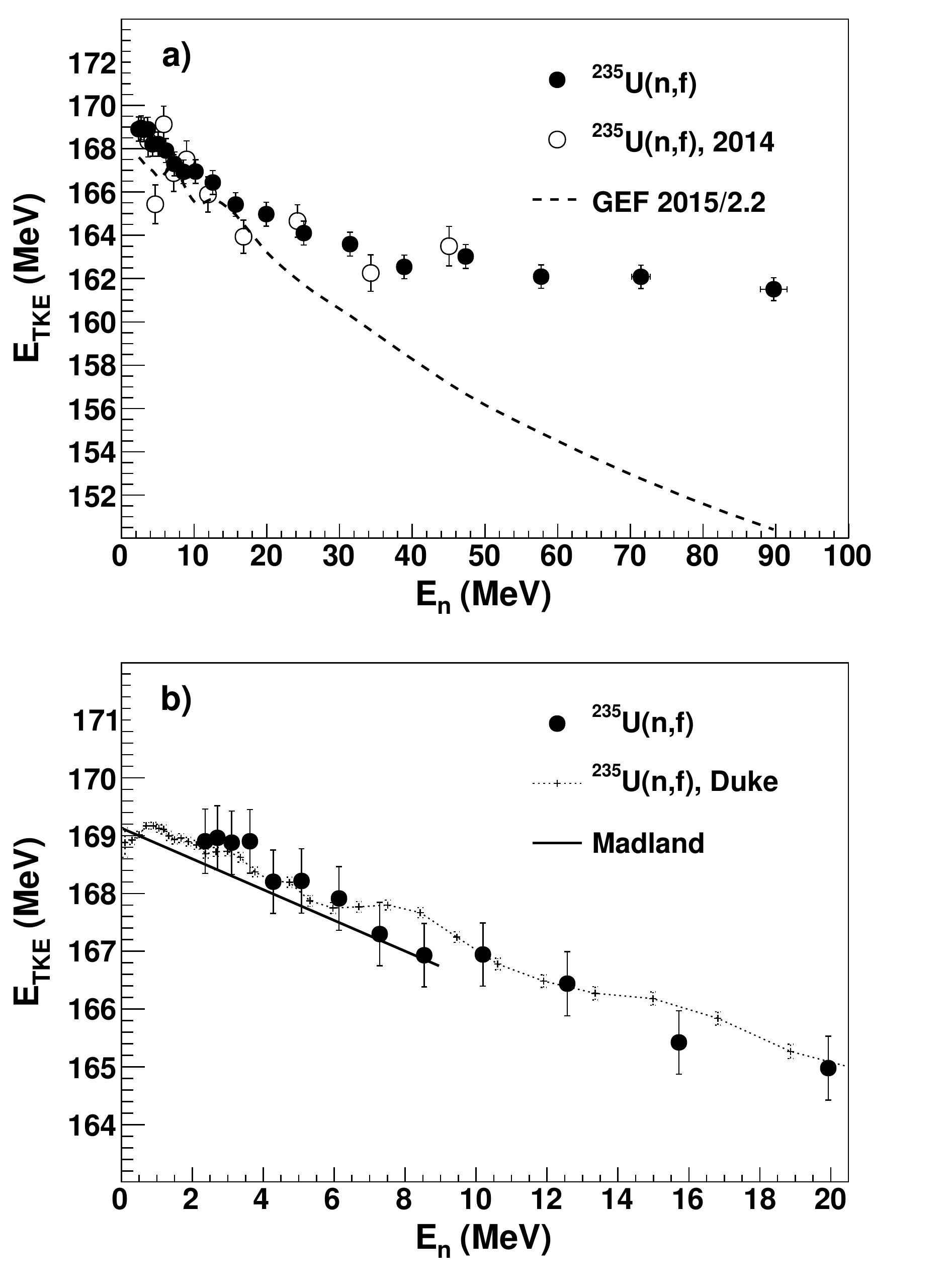}
\caption{Panel a) Post-neutron emission TKE as a function of neutron energy in the $^{235}$U(n,f) reaction for $2<E_n<100$ MeV. Solid symbols represents the current data and the open symbols our previous measurement \cite{yanez14}. The dashed line is the calculation with the GEF code \cite{gef}. Panel b)  Post-neutron emission TKE as a function of neutron energy in the $^{235}$U(n,f) reaction for $2<E_n<20$ MeV. The + symbols connected by dots represent the unpublished relative data from Ref.~\cite{duke15} and the solid line is the fit made by Madland \cite{madland06} to the data of Straede \textit{et.\ al} \cite{straede87} and Meadows \textit{et.\ al} \cite{meadows82}.}
\label{fig_tke}
\end{figure}

\begin{figure}[hbp]
\includegraphics[scale=0.8]{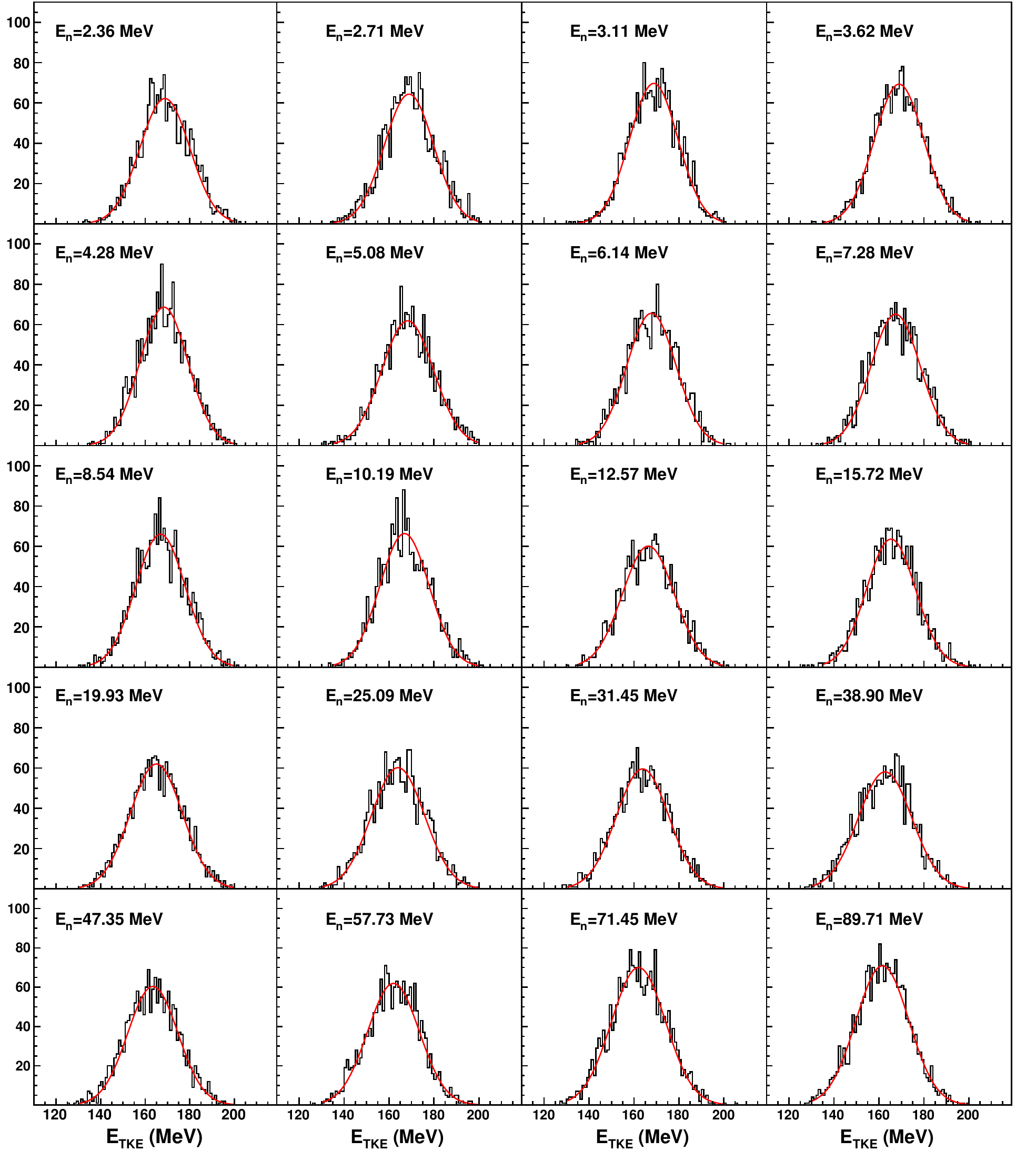}
\caption{Post-neutron emission TKE distributions for each energy bin. The solid red line in each panel represent a fit made with a Gaussian distribution. The limits of each neutron energy bin are listed in Table~\ref{tab1}.}
\label{fig_tkedists}
\end{figure}

In Fig.~\ref{fig_tke} panel a) we show the present TKE measurement of the $^{235}$U(n,f) reaction (solid symbols) as a function of incident neutron energy (see also Table~\ref{tab1}), together with our previous data \cite{yanez14} (open symbols). The dashed line is a calculation with the GEF code using the standard model parameters. In panel b) of Fig.~\ref{fig_tke} we show a subset of the data for $2<E_n<20$ MeV, the unpublished data of Ref~\cite{duke15} and the fit made by Madland \cite{madland06} to the data of Refs.~\cite{straede87,meadows82} (solid line). In Fig.~\ref{fig_tkedists} we show the TKE distributions in each energy bin. The solid line in each panel represent a fit with a Gaussian distribution. The distributions are all nearly Gaussian.

The thermal neutron-induced data measured at Oregon State University was analyzed using the same method and corrections as the fast-neutron induced data, except with fixed $\nu_{tot}=2.43$ \cite{endf11} and $\nu_{pre}=0$. The measured thermal neutron-induced post-neutron emission TKE is $E_{TKE}^{th}=169.8 \pm 0.4$ MeV and $\sigma_{TKE}^{th}=10.18\pm 0.02$ MeV. The pre-fission TKE is calculated from the approximation \cite{madland06},
\begin{equation}
\left<E_{TKE}\right>=\left<E_{TKE}^*\right>\left[ 1-\frac{\nu_{tot}}{2A_{CN}^*}\left(\frac{\left<A_H^*\right>}{\left<A_L^*\right>}+\frac{\left<A_L^*\right>}{\left<A_H^*\right>}\right)\right]
\end{equation}
\noindent
which gives $E_{TKE}^{th,*}=170.7 \pm 0.4$ MeV. This result is consistent with the previous measurements of $171.9\pm1.4$ of Ref.~\cite{schmitt66} and $172.0\pm2.0$ of Ref.~\cite{bennett67}, and in close agreement with the recently recommended value for benchmarking fission theory \cite{bertsch15}. It is important to note that the data in Ref.~\cite{duke15} shown in Fig.~\ref{fig_tke} are relative, scaled with an unpublished theoretical value, whereas the data of the present measurement are absolute and benchmarked against the known value of the TKE in the $^{235}$U(n$_{th}$,f) reaction.

The linear decrease in TKE predicted by Madland for $0 < E_n < 10$ MeV is 0.266 MeV per MeV increase in $E_n$ (see Fig.~\ref{fig_tke}). There has been much speculation about the physical origin of the decrease in TKE, being tentatively attributed to the change of the potential energy surface in deformation space, from producing asymmetric mass splits at low excitation energy to symmetric ones at higher energies \cite{madland06}. Ultimately, shell effects are responsible for the asymmetric component and their fade-out with excitation energy is responsible for the slow but steady growth of the symmetric component as the energy of the incoming neutron increases. It has been pointed out, however, that the asymmetric component may decrease and the symmetric may increase at a much slower pace than would be required for the fall of TKE to be explained solely in terms of the change of the potential energy surface \cite{lestone14}.

To understand the dependence of the TKE in terms of mass symmetric and asymmetric fission, the experimental pre-neutron emission mass distributions have been fit with Gaussian functions representing different fission modes, symmetric and asymmetric, and the TKE has been analyzed in terms of these modes.

Hambsch \textit{et.\ al} \cite{hambsch03} made fits to the mass distributions measured by Straede \textit{et.\ al} \cite{straede87} in the $^{235}$U(n,f) reaction, $E_n\le 5.5$ MeV, to constrain the parameters of a multi-modal fission model based on the Los Alamos model \cite{madland82} to ultimately deduce the prompt neutron multiplicity and energy spectra. Three prominent fission modes were considered; the superlong (SL), standard I (S1) and standard II (S2) modes. The names where coined by Brosa, Grossmann and M\"uller \cite{brosa90} in describing the bifurcations of pre-scission shapes within the shell-corrected liquid-drop model. In the case of U$^{236}$, the asymmetric modes are driven by the shells of doubly magic $^{132}$Sn (standard I) and the deformed neutron shell $N=88$ (standard II), occurring after the saddle point. Therefore, these models are often referred to as scission-point models, to distinguish them from the standard transition state models where fission is determined at the saddle point, and follow the pioneering work of Wilkins, Steinberg and Chasman \cite{wilkins76}. Hambsch \textit{et.\ al} found $A_H^{S1}=134$ and $A_H^{S2}=141$ for the standard I and standard II modes. The symmetric fission mode, was fixed to $A^{SL}=118$, as was its standard deviation $\sigma_{SL}=15$. Incidentally, the same average masses were found in the multi-modal analysis of $^{235}$U(n$_{th}$,f) \cite{hambsch89}.

In an attempt to reproduce this result, we have constructed pre-neutron emission mass distributions with our data for $E_n=$2.5, 3.0, 3.5, 4.0, 4.5, 5.0 and 5.5 MeV in narrow energy bins of $\pm 100$ keV and performed fits with the assumption of three fission modes, one symmetric centered at $A^{SL}=118$, and two asymmetric centered at $A_H^{S1}=134$ and $A_H^{S2}=141$, respectively. In order to reproduce the branching ratios reported by Hambsch \textit{et.\ al}, the complementary fragments $A_L^{S1}=102$ and $A_L^{S2}=95$, and the standard deviations $\sigma_{S1}=3.2$, $\sigma_{S2}=5.5$ had to be fixed as well. Hence, only the normalization parameters were allowed to vary freely. With these restrictions, the branching ratios $b$ of the modes were found to be very similar to those found in Ref.~\cite{hambsch03}, namely, $b_{S1}\sim 0.25$, $b_{S2}\sim 0.75$ and $b_{SL}$ very small but increasing steadily to $\sim$ 0.02 at the highest energy.

In order to extend the mass distribution fits to higher energies ($6<E_n<100$ MeV), beyond first-chance fission, we make the following assumptions: 1) $A_H^{S1}=134$ and $A_H^{S2}=141$ at all energies, 2) $A^{SL}=A^*/2$,  $A_L^{S1}=A^*-A_H^{S1}$ and $A_L^{S2}=A^*-A_H^{S2}$. 3) $\sigma_{S1}=3.2$, $\sigma_{S2}=5.5$ and $\sigma_{SL}=15$. $A^*$ is again the mass of the pre-fission nucleus, $A^*=236-\nu_{pre}(E_n)$. In other words, we assume the extra stability around doubly magic $^{132}$Sn and the deformed neutron shell is preserved for all excitation energies, and that the standard deviations do no significantly depend of the nuclear temperature at the saddle. An attempt to find a tentative temperature dependence of the standard deviations was performed by Straede \textit{et.\ al} \cite{straede87}. We will later argue that the pre-fission excitation energy is rather constant; the initial excitation energy is effectively removed by pre-fission emission of neutrons. Any dependence of the standard deviations with the nuclear temperature at the saddle is therefore modest in the entire energy range.

\begin{figure}[hbp]
\includegraphics[scale=0.8]{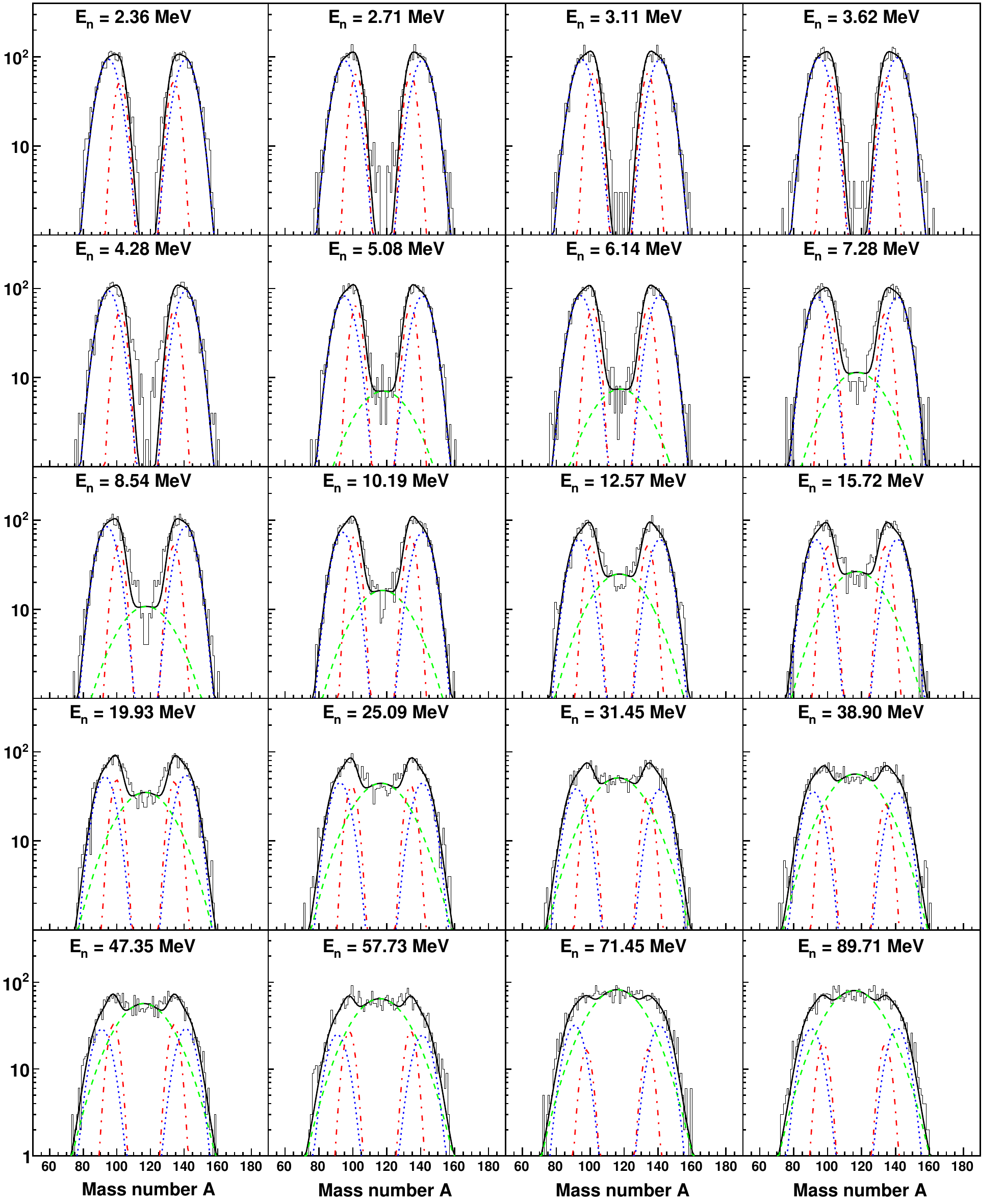}
\caption{Mass distribution in the $^{235}$U(n,f) reaction for $2<E_n<100$ MeV. Solid lines represent the fit with five Gaussian distributions representing the standard I (dot-dashed), standard II (dotted) and superlong (dashed) fission modes. The limits of each neutron energy bin are listed in Table~\ref{tab1}.}
\label{fig_massdists}
\end{figure}

\begin{figure}[hbp]
\includegraphics[scale=0.5]{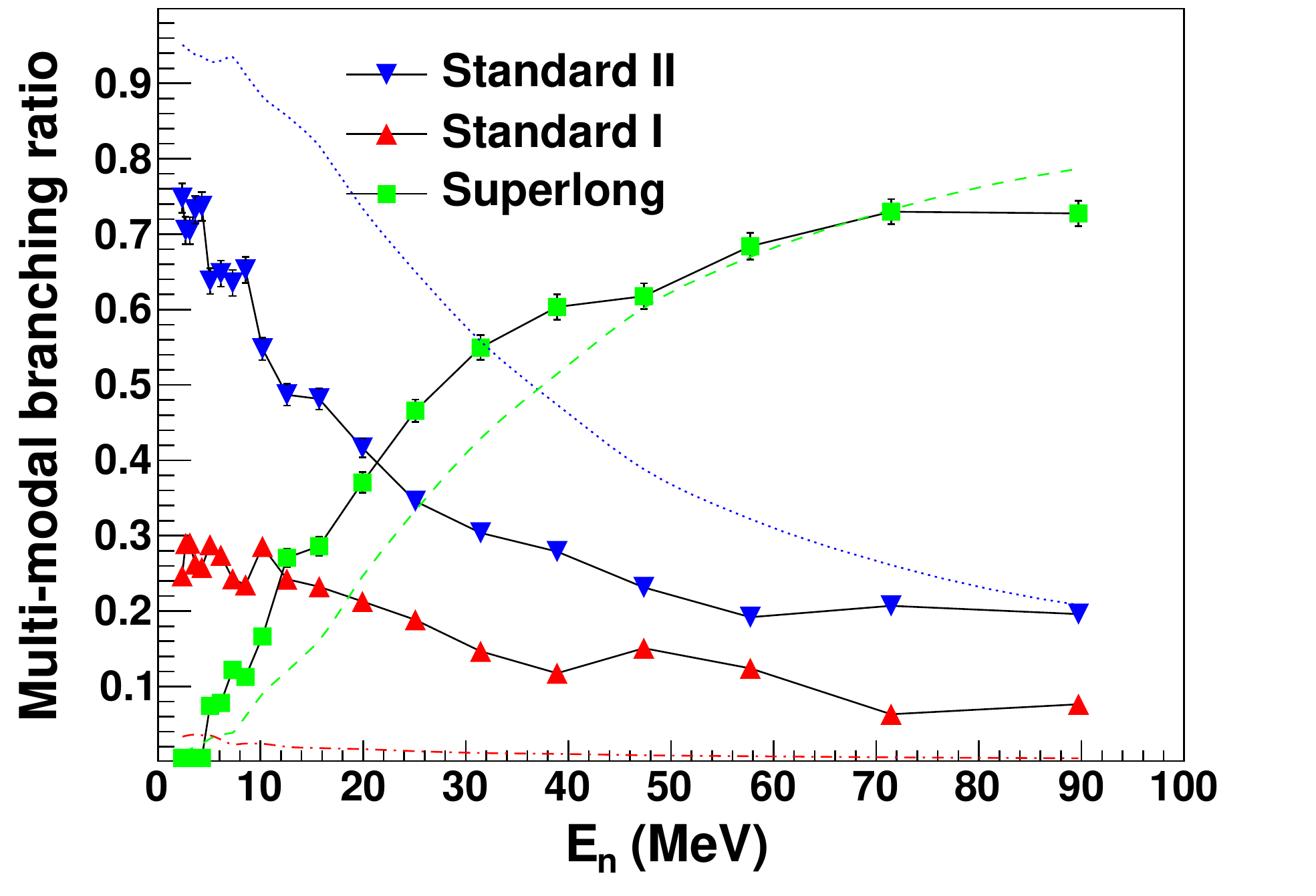}
\caption{Experimental multi-modal branching ratios as a function of incident neutron energy in the $^{235}$U(n,f) reaction. The broken lines represent the calculation with the GEF code \cite{gef}; standard I (dot-dashed), standard II (dotted) and superlong (dashed).}
\label{fig_branches}
\end{figure}

In Fig.~\ref{fig_massdists} we show the mass distributions in the same energy bins as the TKE distributions. The solid lines represent the fit, whereas the broken lines represent the contributions from the standard I (dot-dashed), standard II (dotted) and superlong (dashed) modes. In Fig.~\ref{fig_branches} we show the branching ratios as a function of incident neutron energy (Table~\ref{tab2} lists the data.) At the highest energy, the asymmetric mass modes persist with $\sim 30$\%. 

\begin{figure}[hbp]
\includegraphics[scale=0.5]{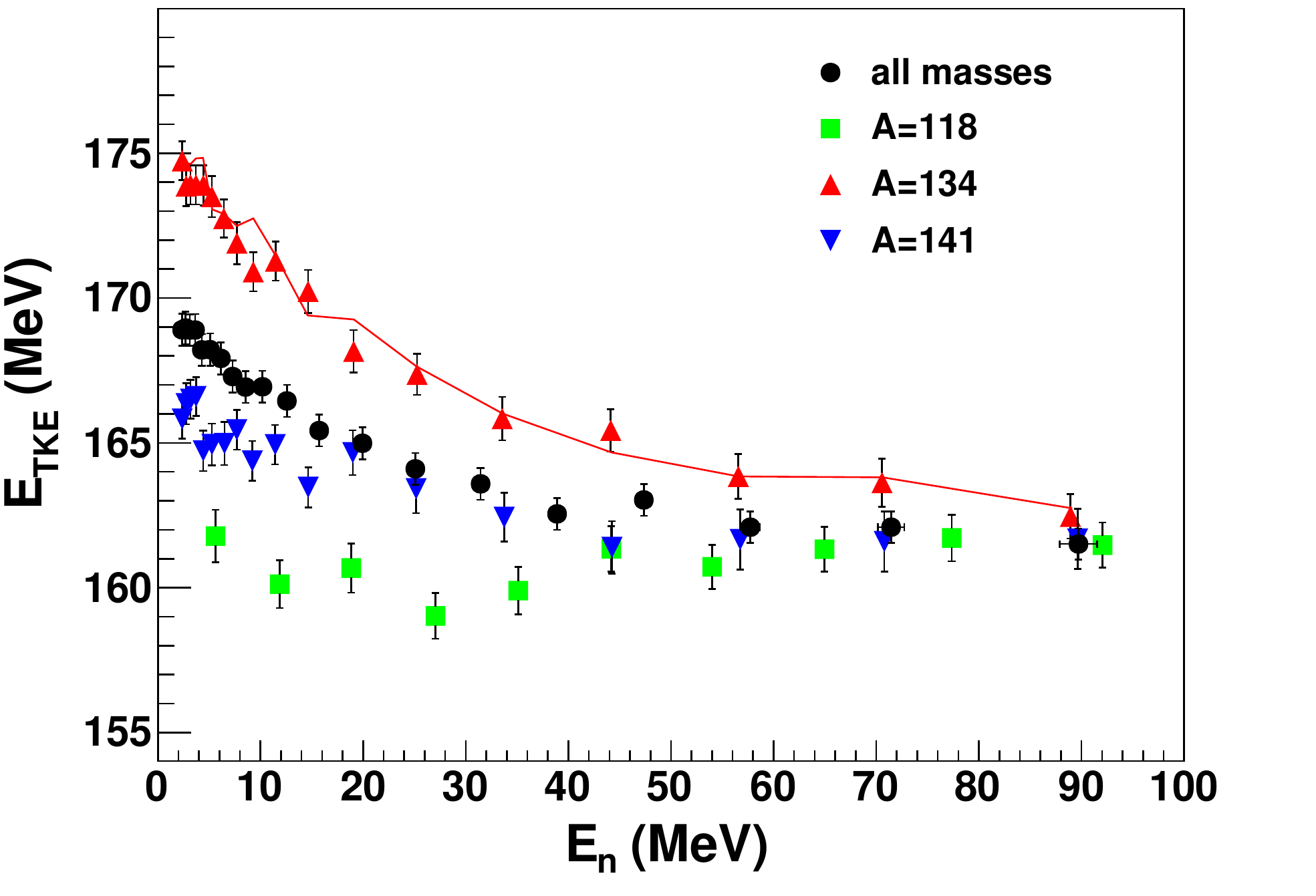}
\caption{TKE as a function of neutron energy around the average mass of the standard I, standard II and superlong modes in the $^{235}$U(n,f) reaction. The solid (red) line is a fit around $A=134$ assuming the TKE is given by a superposition of modes (see text for details).}
\label{fig_tkea}
\end{figure}

\begin{figure}[hbp]
\includegraphics[scale=0.5]{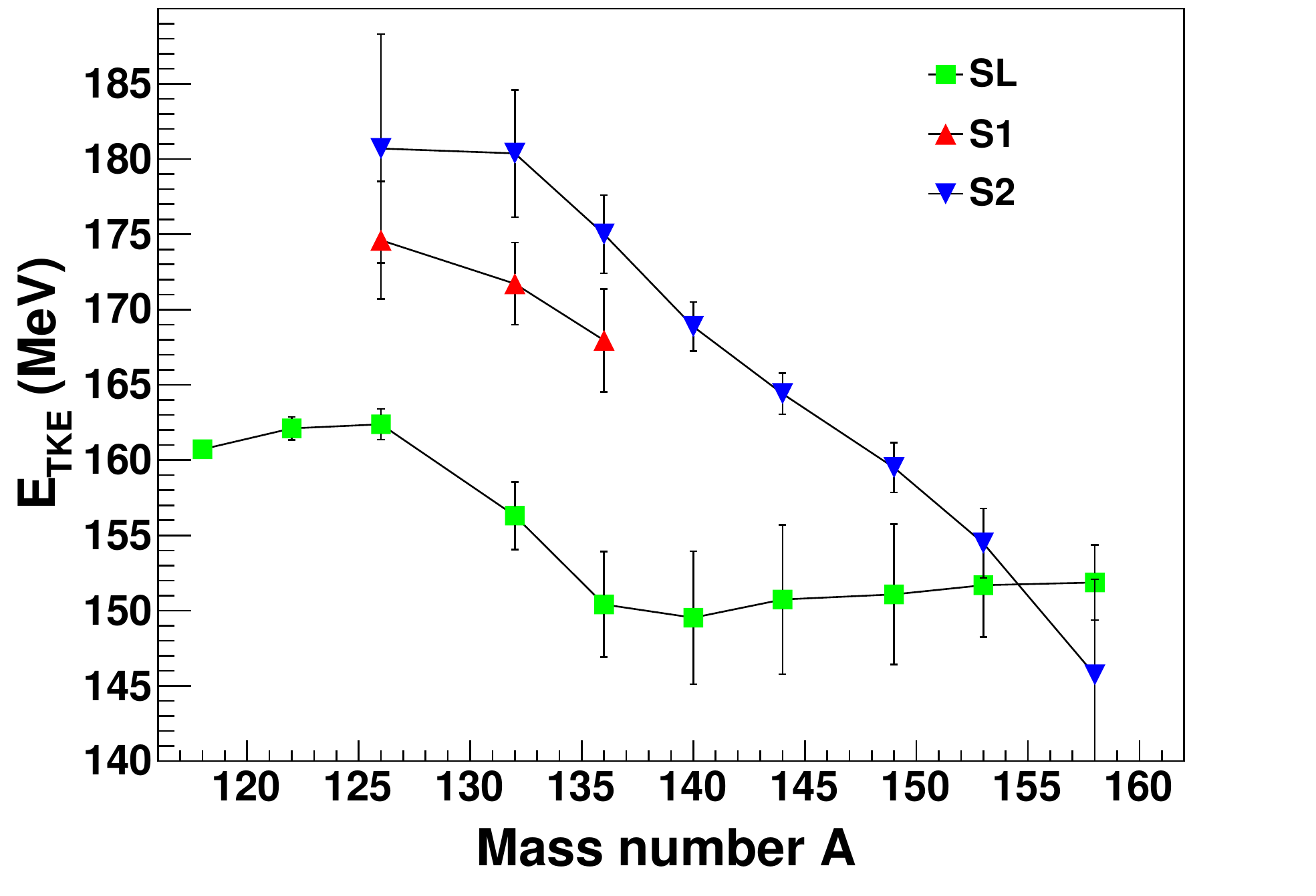}
\caption{Multi-modal TKE as a function of mass in the $^{235}$U(n,f) reaction.}
\label{fig_tkemodes}
\end{figure}

In Fig.~\ref{fig_tkea} we show the TKE by making cuts in the mass distribution around the average mass of the superlong ($A=118$), standard I ($A=134$) and standard II ($A=141$) modes. The optimum width of the cut, given the experimental mass resolution, was determined by the Monte Carlo simulations and varies with mass. It may at first seem as if the standard I mode is associated with a higher TKE compared to the standard II and superlong modes, but one quickly realizes the TKE in each mass cut stems from contributions from each mode in different proportions. To separate the TKE associated with the modes, the data is fit assuming the TKE is a linear superposition of the average TKE of each mode,
\begin{equation}
E_{TKE}=b_{S1}|_A\left<E_{S1}\right>+b_{S2}|_A\left<E_{S2}\right>+b_{SL}|_A\left<E_{SL}\right>
\label{eq4}
\end{equation}
\noindent
where $b_{mode}|_A$ represents the branching ratio of the mode around mass $A$ and is evaluated by integration of the multi-modal fits shown in Fig.~\ref{fig_massdists}. $\left<E_{mode}\right>$ is the average TKE of the mode and is by assumption independent of incident neutron energy. The solid line in Fig.~\ref{fig_tkea} is the fit around $A=134$ and yields $\left<E_{S1}\right>=172.2\pm2.7$ MeV, $\left<E_{S2}\right>=174.0\pm2.6$ MeV and $\left<E_{SL}\right>=149.0\pm3.1$ MeV. The reduced $\chi^2$ of the fit is 0.48 with 15 degrees of freedom. In Fig.~\ref{fig_tkemodes} we show the average TKE for each mode as a function of mass of the heavy fragment. It appears the TKE of the asymmetric modes are rather similar and decrease steeply with increasing mass asymmetry, whereas the TKE of the symmetric mode is rather constant in comparison. The average TKE of the most prominent fission modes seem to depend on the mass split but not on incident neutron energy. Being the underlying assumption of Eq.~\ref{eq4}, this suggests the decrease in the overall TKE shown in Fig.~\ref{fig_tke} and expressed as a linear function by Madland \cite{madland06} is indeed a consequence of the growth of symmetric and the demise of asymmetric fission as shell effects are washed out with increasing excitation energy. The average reduced $\chi^2$ of all fits is $\sim 0.92$.

\subsection{The width of the total kinetic energy distribution}

\begin{figure}[hbp]
\includegraphics[scale=0.5]{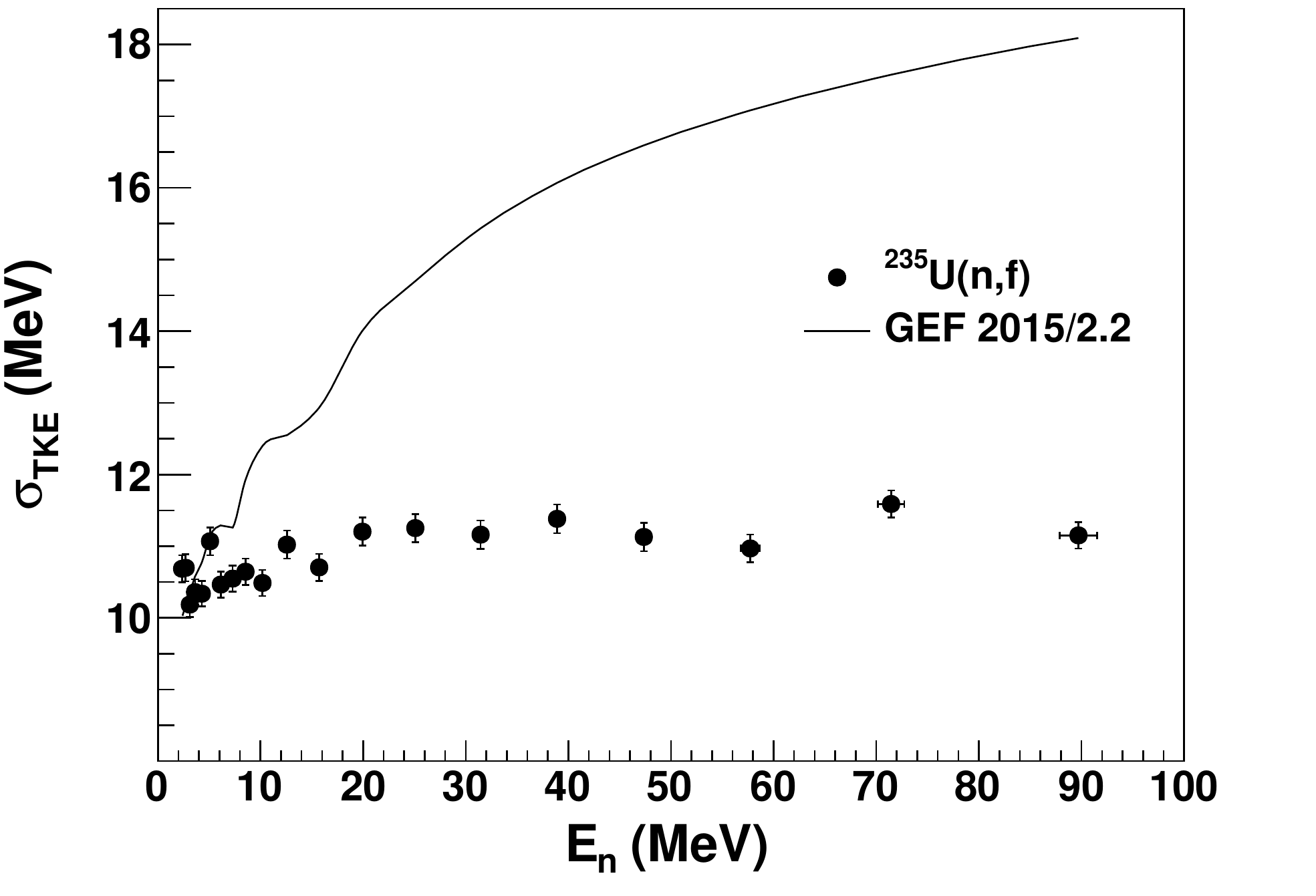}
\caption{Post-neutron emission $\sigma_{TKE}$ in the $^{235}$U(n,f) reaction. Solid circles represent the current data. The solid line is the calculation with the GEF code \cite{gef}.}
\label{fig_var}
\end{figure}

\begin{figure}[hbp]
\includegraphics[scale=0.5]{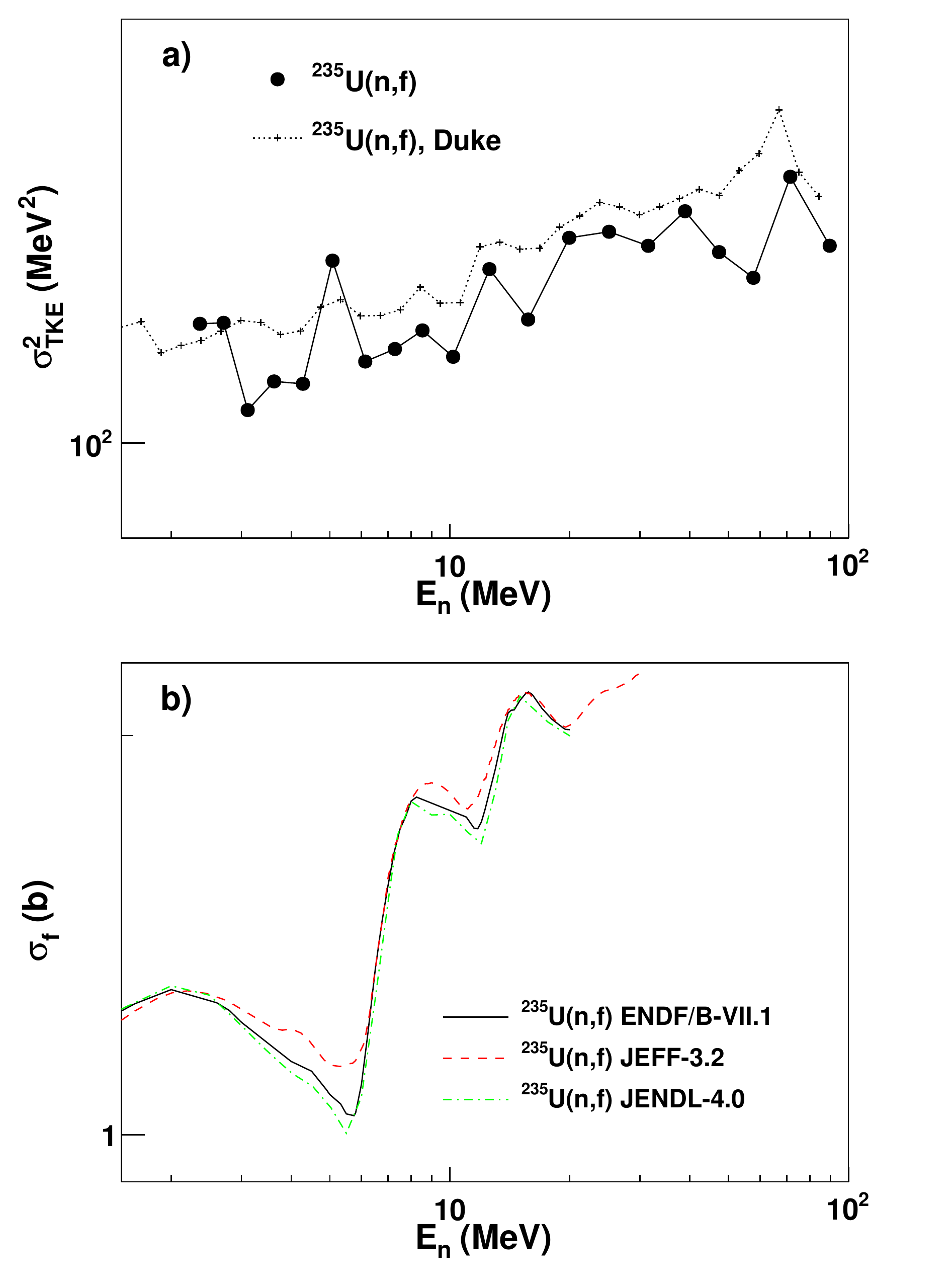}
\caption{Panel a) Post-neutron emission variance in the $^{235}$U(n,f) reaction in log-log scale. The + symbols connected by dots represent the unpublished data from Ref.~\cite{duke15}. Panel b) Fission cross section in the $^{235}$U(n,f) reaction. Data taken from the ENDF/B-VII.1 \cite{endf11}, JEFF-3.2 and JENDL-4.0 \cite{jendl11} libraries.}
\label{fig_var2}
\end{figure}

The dependence of the width of the TKE distribution on excitation energy has been found to increase in $^{235}$U(p,f), $E_p=8-13$ MeV \cite{ferguson73} and $^{235}$U($\alpha$,f), $E_\alpha=20-80$ MeV \cite{zaika85}. In our previous measurement reported in Ref.~\cite{yanez14} we found no evidence of an increase of $\sigma_{TKE}$ with neutron energy in the $^{235}$U(n,f) reaction, $E_n=20-50$ MeV. We confirm this unexpected result. In Fig.~\ref{fig_var} we show the overall $\sigma_{TKE}$ as a function of incident neutron energy. $\sigma_{TKE}$ seems to increase gently for $E_n<20$ MeV and is essentially constant in the range $20<E_n<100$ MeV. The code GEF calculates a steeper increase with neutron energy (solid line). Many models of fission associate the $\sigma_{TKE}$ with fluctuations in both the separation distance and kinetic energy of the fission fragments at scission \cite{wilkins76,grossmann88}. Fluctuations should increase with excitation energy, which is why the measured constancy of $\sigma_{TKE}$ is so puzzling. Interestingly, the $\sigma_{TKE}$ in Ref.~\cite{duke15} for $0<E_n<20$ MeV show statistically significant variations that are attributed to changes in the deformation of the fissioning nucleus around the multi-chance fission thresholds \cite{duke15}. Such variations are also visible in our measurement, at similar neutron energies. To make these variations more visible, in Fig.~\ref{fig_var2} panel a) we show the TKE variance in log-log scale together with the data of Ref.~\cite{duke15} (uncorrected for $\sigma_{inst}$.) The sudden increase in variance at specific energies seem related to the onset of a new fission chance, similar to the variations observed in the cross sections, shown in Fig.~\ref{fig_var2} panel b).

\section{Discussion}
\label{discussion}

\begin{figure}[hbp]
\includegraphics[scale=0.5]{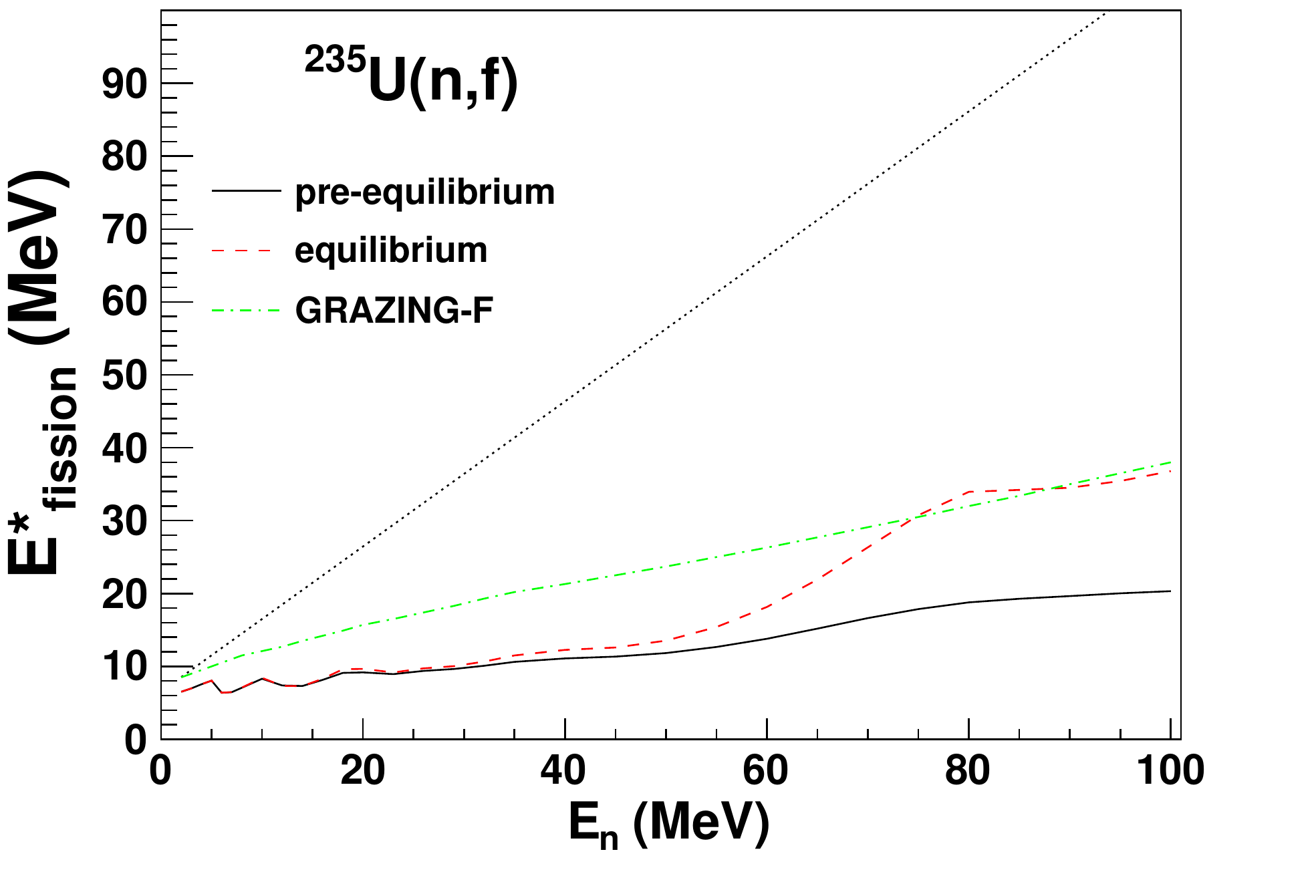}
\caption{Calculated pre-fission excitation energy as a function of incident neutron energy. The dotted line represents the initial excitation energy, the solid line the calculation with TALYS including pre-equilibrium emission, the dashed line the calculation with TALYS with only equilibrium emission and the dot-dashed line is a calculation with the fission model of GRAZING-F \cite{yanez15}.}
\label{fig_exci}
\end{figure}

\begin{figure}[hbp]
\includegraphics[scale=0.5]{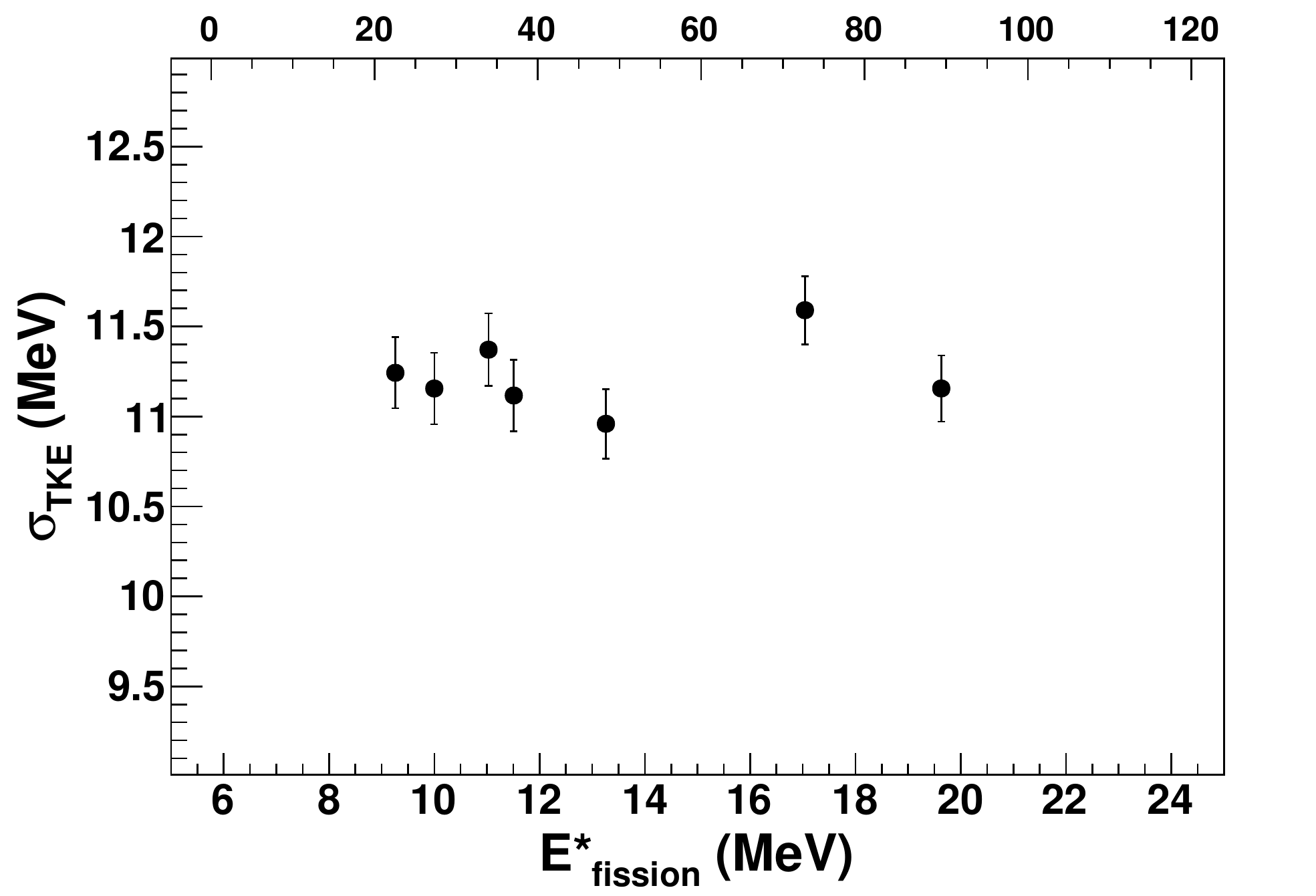}
\caption{Measured post-neutron emission $\sigma_{TKE}$ in the $^{235}$U(n,f) reaction and $E_n>20$ MeV as a function of the pre-fission excitation energy. The top x-axis shows the approximate scale of the incident neutron energy $E_n$.}
\label{fig_varE}
\end{figure}

Early studies of the standard deviation of TKE distributions concluded the dispersion of TKE was not entirely due to neutron emission and was inherent to the fission process itself \cite{unik64}. Near the same time, liquid drop models suggested $\sigma_{TKE}$ was temperature dependent \cite{nix63}. Later models considered the physical processes taking place from the saddle to the scission points, in particular, the partial transformation of the potential energy into kinetic energy and its dissipation due to friction \cite{davies77,grossmann88}. The weak increase of $\sigma_{TKE}$ shown in Fig.~\ref{fig_var} suggests the temperature dependence may not be related to the temperature associated with the initial excitation energy of the fissioning nucleus. Since excitation energy is being dissipated by evaporation and friction while the nucleus is progressing towards scission, the temperature $\sigma_{TKE}$ is sensitive to may be the temperature at the saddle or scission points, or somewhere in between. We have calculated the average pre-fission excitation energy using the TALYS \cite{talys} code and the model described in Ref.~\cite{yanez15}. In the TALYS calculations, we also consider pre-equilibrium emission as a possible reaction mechanism. In Fig.~\ref{fig_exci} we show the calculated average pre-fission excitation energy. The dotted line represents the initial excitation energy, $E^*=Q+E_{c.m.}$. Pre-fission neutron emission, emitted either before or after statistical equilibrium is attained, reduces the initial excitation energy prior to fission. The solid line represents the TALYS calculation in which pre-equilibrium emission is considered, the dashed line the calculation in which only emission from a fully equilibrated nucleus is considered, and the dot-dashed line represents the calculation with a fission model commonly used in heavy element research \cite{yanez15}. Comparison between the TALYS calculations reveal that the predicted threshold for pre-equilibrium processes is $E_n \sim 50$ MeV for this system. A pre-equilibrium particle is emitted, by definition, from a non-equilibrated source, and removes a larger amount of excitation energy from the excited nucleus compared to equilibrated evaporation. Although the pre-fission neutron multiplicity in the two TALYS calculations remain fairly similar, once the pre-equilibrium threshold is reached, the excitation energy prior to fission is removed to a greater extent by pre-equilibrium processes. These particles are preferentially emitted in the direction of the incoming neutron and their kinetic energy distributions have a Maxwellian shape with a slope consistent with a very high nuclear temperature. This high temperature is apparent, as it reflects a situation in which the energy brought in by the incoming neutron has not yet been thermalized. In Fig.~\ref{fig_varE} we show $\sigma_{TKE}$ plotted as a function of the pre-fission excitation energy (TALYS calculation including pre-equilibrium emission) in the $^{235}$U(n,f) reaction for $E_n>20$ MeV. The top x-axis shows the approximate scale of the incoming neutron energy. The experimental $\sigma_{TKE}$ has no significant dependence on the excitation energy of the nucleus just before fission. If $\sigma_{TKE}$ has a weak dependence on excitation energy, what does it depend on? 

\begin{figure}[hbp]
\includegraphics[scale=0.5]{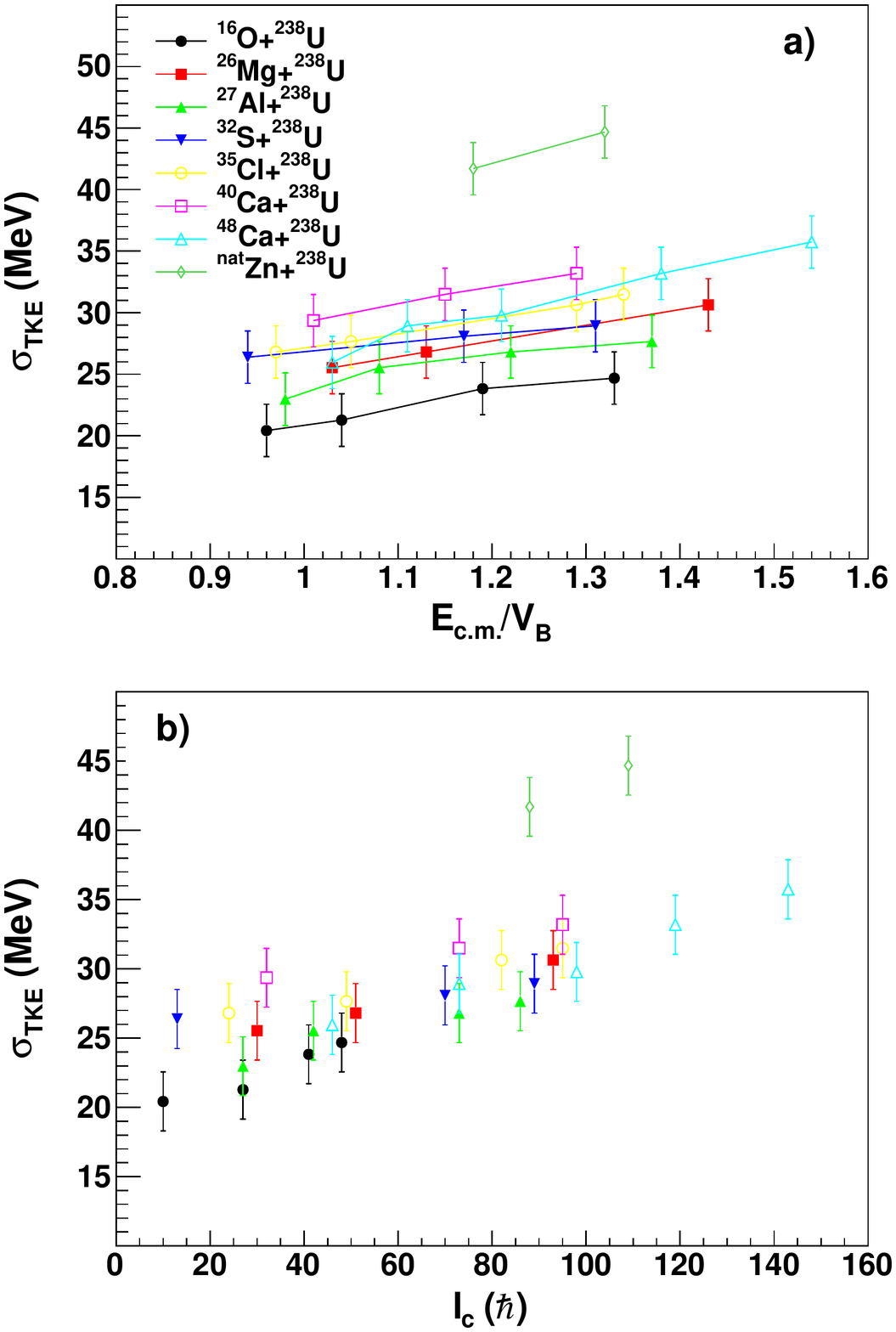}
\caption{Standard deviation of TKE distributions in symmetric fission in $^{238}$U + $^{16}$O, $^{24}$Mg, $^{27}$Al, $^{32}$S, $^{35}$Cl, $^{40}$Ca, $^{48}$Ca, $^{nat}$Zn reactions \cite{shen87}. In panel a) the data is plotted as a function of $E_{c.m.}/V_B$, in panel b) $\sigma_{TKE}$ is plotted as a function of $\ell_{c}$. }
\label{fig_back}
\end{figure}

In capture-fission reactions involving light and medium-size nuclei the experimental TKE distributions are much broader than those found in particle-induced (n, p, d, t, $\alpha$, etc.) fission reactions. In Fig.~\ref{fig_back} we show the widths of TKE distributions measured in $^{16}$O, $^{24}$Mg, $^{27}$Al, $^{32}$S, $^{35}$Cl, $^{40}$Ca, $^{48}$Ca and $^{nat}$Zn targets with $^{238}$U projectiles \cite{shen87}. In panel a) the $\sigma_{TKE}$ is plotted as a function of $E_{c.m.}/V_B$, where $V_B$ is the interaction barrier. The data may imply that $\sigma_{TKE}$ strongly correlates with excitation energy, which is the accepted interpretation, but one has to remember that ion-induced reactions also transfer substantial angular momentum to the fissioning nuclei. In capture-fission reactions, angular momentum is transferred in the range [$0-\ell_{c}$] $\hbar$, where $\ell_{c}$ is the maximum angular momentum leading to capture. Since particle-induced fission reactions seem to suggest the excitation energy dependence of $\sigma_{TKE}$ is rather weak, and these reactions transfer insignificant angular momentum, perhaps angular momentum is the quantity that drives the widths of TKE distributions. In panel b) of Fig.~\ref{fig_back} we show $\sigma_{TKE}$ measured in Ref.~\cite{shen87}, plotted as a function of $\ell_{c}$ deduced from the capture cross sections and the sharp cut-off assumption. There appears to be a distinct relation, perhaps driven by the changes in the potential energy surface induced by large angular momentum, which affects the fission barriers, the location of the scission points, the widths of the necks, and hence the overall paths the fissioning system takes towards scission. In other words, the span of potential energy surfaces available due to the larger range of angular momentum are perhaps such that it leads to a wider set of scission configuration ensembles, in particular, sets of scission configurations in which the nascent fragments are closer together or farther apart, that are not readily accessible to particle-induced fission reactions. This picture could explain why $\sigma_{TKE}$ in $^{235}$U(n,f) show no apparent excitation energy dependence for $E_n>20$ MeV. It has been assumed that TKE variances are solely driven by fluctuations, when perhaps they are also driven by the dependence of the potential energy surface on angular momentum. This picture explains simultaneously the data in Fig.~\ref{fig_varE} and Fig.~\ref{fig_back}. The fluctuation hypothesis can only explain the data as plotted in Fig.~\ref{fig_back} panel a). In events where there is a large angular momentum transfer, some of the energy is tied up in rotational energy, reducing the available excitation energy that perhaps drives the contribution to $\sigma_{TKE}$ from fluctuations, making them even less apparent. In the $^{235}$U(n,f) reaction at $E_n=20$ MeV the average angular momentum of the compound nucleus calculated using optical model transmission coefficients (from TALYS) is 4.7 $\hbar$, whereas at $E_n=100$ MeV it is 9.1 $\hbar$. A linear fit to the data in Fig.~\ref{fig_back} panel b) reveals that $\sigma_{TKE}$ may only change by one unit every $\sim 10\hbar$ of transferred angular momentum. Hence, we may suspect the effect on $\sigma_{TKE}$ due to the transferred angular momentum may be difficult to discern in neutron- and particle-induced reactions \cite{back79}.

The possible dependence of $\sigma_{TKE}$ on angular momentum has been studied in the past. Two seminal measurements made by Plasil \textit{et al.} \cite{plasil66} and Unik \textit{et al.} \cite{unik69} studied TKE and its width in reactions leading to the same compound nucleus and excitation energies using several suitable combinations of projectile and target. In Ref.~\cite{plasil66}, the reactions $^{12}$C+$^{174}$Yb and $^{16}$O+$^{170}$Er populated the nuclide $^{186}$Os at $E^*=88$ and $102$ MeV, respectively, whereas in Ref.~\cite{unik69} the reactions p+$^{209}$Bi, $\alpha$+$^{206}$Pb populated nuclide $^{210}$Po at $E^*=31, 44, 57$ MeV and $^{12}$C+$^{198}$Pb at $E^*=58$ MeV. Both studies concluded that the TKE is insensitive to and the widths sensitive to the excitation energy. However, the data of the former study with heavier projectiles suggests the width of the TKE distribution is also dependent on angular momentum, whereas the latter showed a complete insensitivity. 

\section{Conclusions}
\label{conclusions}

The findings in this work are, a) the decrease of TKE with incident neutron energy due to the increase of symmetric fission, and b) the widths of TKE distributions are weakly dependent on excitation energy, and may have a much stronger angular momentum dependence. The variance of TKE show variations related to the onset of fission chances, similar to cross sections.

The multi-modal fission theory of Brosa, based on the shell-corrected liquid-drop model, postulates the existence of two asymmetric modes, believed to be a bifurcation of the asymmetric mode occurring after the saddle point and driven by the shells of doubly magic $^{132}$Sn (standard I) and the deformed neutron shell $N=88$ (standard II). However, fitting the fission mass distributions with Gaussian functions representing the two asymmetric and one symmetric modes does not result in the unambiguous determination of the individual means and variances of the modes. The fits have to be restricted to fixed means and widths for the minimization procedure to numerically converge to physically meaningful values. The average modal TKE does not seem to depend on neutron energy and the TKE decrease results from the change in the relative contributions of symmetric and asymmetric fission as the incident neutron energy increases. The near constant widths of the TKE distribution for $E_n>20$ MeV suggests it is weakly dependent on the initial or pre-fission excitation energy. Given that the excitation energy available at the moment of fission is a largely unknown quantity and can currently only be estimated by fission model calculations, there is strong incentive for measuring the multiplicities and angular distributions of prompt neutrons in particle-induced fission reactions. Such studies shall discern not only the total multiplicity, but the pre- and post-fission, and pre-equilibrium components at higher energies, with which the pre-fission excitation energy can be determined.

The GEF code \cite{gef} is fairly able to account for the mass distributions, as evidenced by the rather good estimation of the branching ratios shown in Fig.~\ref{fig_branches}. GEF is also able to reproduce the measured total prompt neutron multiplicities (Fig.~\ref{fig_nutot}). However, GEF does not account for the TKE means and widths (Fig.~\ref{fig_tke} and Fig.~\ref{fig_var}.)

\appendix

\section{Monte Carlo detector response simulation}

\label{appA}

To understand the inevitable experimental setup biases in the data, a detector response simulation was done using the Monte Carlo method. A simulated event was generated by first sampling a position in the $45^\circ$ tilted target with a flux of incoming particles distributed with a Gaussian of 1.0 cm FWHM perpendicular to the beam direction. Then an isotropic direction in space was generated with the Marsaglia method \cite{marsaglia}. By using ray-tracing techniques, the chosen direction is tested for interception with a detector in either array. If the direction generates a hit, the opposite direction is randomly sampled with a cone of aperture $\alpha$ relative the cone axis (identical with the chosen direction). This randomization of the complementary direction intends to mimic the deflections given to the fission fragments by the recoil in the emission of neutrons and angular straggling due mainly to atomic collision in the target and backing materials. The randomized direction is tested again for a hit in a detector in the opposite array. If a double hit is generated, the $Z$, $A$ (mass and charge conservation are assured), and energies of two fragments is sampled with Gaussian distributions of given means and variances. The sampled energies are further degraded by calculating the energy loss using range tables \cite{northcliffe} from the location in the target where the decay happens, the depth of which is sampled randomly, to the exit of the target and/or backing material along the chosen direction. The degraded energies are further sampled with a Gaussian distribution of width $\Delta E$, to mimic the intrinsic energy resolution of the detectors, as is the calculated time-of-flight (TOF) of the fragments with a Gaussian of width $\Delta t$, to mimic the intrinsic time resolution of the detectors. The inverse of the Schmitt function is calculated with the final energies, rendering the simulated channel number in the corresponding ADC\@. All the parameters of the simulated event are saved in a file, including the identity of the detectors that hit, $A$ and $Z$ of the fragments, every sampled energy, direction of cosines, TOF and ADC channel. The file of simulated events is then analyzed as if it were experimental data, using only the simulated information that is readily available in the experiment; the ADC channel and the TOF of the coincident fragments, and then compared to the initially generated values.

The intrinsic detector characteristics, the energy and time resolutions, used as input in the simulations, were estimated by measurement. A new PIN-diode detector has a typical $\alpha$-particle energy resolution of $\sim 17$ keV FWHM for the $5805$ keV line of $^{244}$Cm. The time resolution was measured by detecting coincident $^{252}$Cf fission fragments from the thin source made by evaporating 5 $\mu$l of a 25 nCi/$\mu$l solution on a 100 $\mu$g/cm$^2$ C foil and had an area of 5.4 mm$^2$. The source was ``sandwiched'' between two detectors to minimize the flight distance and optimize the simultaneity of the events. The measured time resolution is $\Delta t \sim 200$ ps for the Hamamatsu PIN-diodes model S3590-09.

There are readily two methods with which the data can be analyzed; by using the energy and time signals ($E\Delta t$ method) or only the energy signals ($2E$ method). The $E\Delta t$ method solves the kinematical equations iteratively by adjusting the masses calculated using the time difference between the fragments. Only mass conservation is assumed. The $2E$ method solves the kinematical equations iteratively by adjusting the masses calculated assuming momentum and mass conservation, as thoroughly reviewed in section~\ref{analysis}. With the help of the simulations it was determined that, given the geometry of the setup, dictated by the neutron beam intensity and the length of the bombardment, the $2E$ method is preferred over the $E\Delta t$ method. If it is assumed the typical energy resolution of the detectors is 0.3\% (FWHM), the resulting mass resolution is $\Delta m=7$ u, whereas with the $E\Delta t$ method, assuming additionally a time resolution of 200 ps, the resulting mass resolution is $\Delta m \sim 25$ u. In terms of $\Delta m$, the $E\Delta t$ method becomes better than the $2E$ method if the distance to the detector arrays is increased by a factor of 5, at which point the solid angle subtended has decreased by a factor of 25. Hence, the $E\Delta t$ is the method that should be used if mass resolution is of crucial importance. The mass resolution obtained with the $2E$ method is solely dependent on the energy resolution of the detectors, whereas the  $E\Delta t$ depends additionally on $\Delta t$ and the detector distance. In the analysis of the present data we have used the $2E$ method.

\begin{acknowledgments}
The authors wish to thank Dr. F. Tovesson for his support, and the reactor staff at the Oregon State University TRIGA Reactor for providing the thermal neutron beam.

This work was supported in part by the Director, Office of Energy Research, Division of Nuclear Physics of the Office of High Energy and Nuclear Physics of the U.S. Department of Energy under Grant DE-FG06-97ER41026 and DE-SC0014380, the U.S. Dept.\ of Energy, NNSA, under Grant DE-NA0002926, and the U.S. Department of Energy, Los Alamos National Security, LLC under contract DE-AC52-06NA25396.

\end{acknowledgments}

\bibliography{TKE-U}

\begin{table}[tbp]
\caption{\label{tab1}Post-neutron emission TKE and $\sigma^2_{TKE}$ as a function of the incident neutron energy in the $^{235}$U(n,f) reaction. The neutron energy bin limits are shown inside square brackets in the first column. The neutron energy is the geometrical mean and the neutron energy error is estimated using the width of the photo-fission peak. The last column is the number of events $N$ in the bin.}
\begin{tabular}{lcccc}
\hline
\hline
& $E_n$ (MeV) & $E_{TKE}$ (MeV) & $\sigma^2_{TKE}$ (MeV$^2$) & $N$ \\
\hline
$[2.16-2.52]$ & $2.36 \pm 0.01$ & $168.9 \pm 0.6$ & $114.2 \pm 0.1$ & 1807 \\
$[2.52-2.90]$ & $2.71 \pm 0.01$ & $169.0 \pm 0.6$ & $114.5 \pm 0.1$ & 1854 \\
$[2.90-3.35]$ & $3.11 \pm 0.01$ & $168.9 \pm 0.5$ & $103.7 \pm 0.1$ & 1924 \\
$[3.35-3.93]$ & $3.62 \pm 0.01$ & $168.9 \pm 0.5$ & $107.3 \pm 0.1$ & 1936 \\
$[3.93-4.65]$ & $4.28 \pm 0.02$ & $168.2 \pm 0.5$ & $106.9 \pm 0.1$ & 1922 \\
$[4.65-5.55]$ & $5.08 \pm 0.02$ & $168.2 \pm 0.6$ & $122.5 \pm 0.1$ & 1841 \\
$[5.55-6.72]$ & $6.14 \pm 0.03$ & $167.9 \pm 0.6$ & $109.5 \pm 0.1$ & 1866 \\
$[6.72-7.85]$ & $7.28 \pm 0.04$ & $167.3 \pm 0.5$ & $111.3 \pm 0.1$ & 1860 \\
$[7.85-9.26]$ & $8.54 \pm 0.05$ & $166.9 \pm 0.5$ & $113.3 \pm 0.1$ & 1898 \\
$[9.26-11.25]$ & $10.19 \pm 0.07$ & $166.9 \pm 0.5$ & $110.0 \pm 0.1$ & 1893 \\
$[11.25-14.00]$ & $12.57 \pm 0.09$ & $166.4 \pm 0.6$ & $121.5 \pm 0.1$ & 1789 \\
$[14.00-17.65]$ & $15.72 \pm 0.13$ & $165.4 \pm 0.5$ & $114.6 \pm 0.1$ & 1813 \\
$[17.65-22.40]$ & $19.93 \pm 0.18$ & $165.0 \pm 0.6$ & $125.5 \pm 0.1$ & 1852 \\
$[22.40-28.10]$ & $25.09 \pm 0.26$ & $164.1 \pm 0.5$ & $126.6 \pm 0.1$ & 1843 \\
$[28.10-35.20]$ & $31.45 \pm 0.37$ & $163.6 \pm 0.6$ & $124.6 \pm 0.1$ & 1780 \\
$[35.20-42.90]$ & $38.90 \pm 0.50$ & $162.5 \pm 0.5$ & $129.5 \pm 0.1$ & 1800 \\
$[42.90-52.10]$ & $47.35 \pm 0.68$ & $163.0 \pm 0.5$ & $123.9 \pm 0.1$ & 1782 \\
$[52.10-63.40]$ & $57.73 \pm 0.92$ & $162.1 \pm 0.5$ & $120.3 \pm 0.1$ & 1833 \\
$[63.40-80.00]$ & $71.45 \pm 1.28$ & $162.1 \pm 0.5$ & $134.3 \pm 0.1$ & 2173 \\
$[80.00-100.00]$ & $89.71 \pm 1.83$ & $161.5 \pm 0.5$ & $124.3 \pm 0.1$ & 2129 \\
\hline
\hline
\end{tabular}
\end{table}

\begin{table}[tbp]
\caption{\label{tab2}Experimental branching ratios of the standard II, standard I and superlong fission modes as a function of neutron energy in the $^{235}$U(n,f) reaction. The limits of each neutron energy bin are listed in Table~\ref{tab1}.}
\begin{tabular}{lccc}
\hline
\hline
$E_n$ (MeV) & $b_{S2}$ & $b_{S1}$ & $b_{SL}$ \\
\hline
$2.36 \pm 0.01$ & $0.75 \pm 0.02$ & $0.25 \pm 0.01$ & $0.01 \pm 0.00$ \\
$2.71 \pm 0.01$ & $0.71 \pm 0.02$ & $0.29 \pm 0.01$ & $0.01 \pm 0.00$ \\
$3.11 \pm 0.01$ & $0.70 \pm 0.02$ & $0.29 \pm 0.01$ & $0.01 \pm 0.00$ \\
$3.62 \pm 0.01$ & $0.73 \pm 0.02$ & $0.26 \pm 0.01$ & $0.01 \pm 0.00$ \\
$4.28 \pm 0.02$ & $0.74 \pm 0.02$ & $0.26 \pm 0.01$ & $0.01 \pm 0.00$ \\
$5.08 \pm 0.02$ & $0.64 \pm 0.02$ & $0.29 \pm 0.01$ & $0.07 \pm 0.01$ \\
$6.14 \pm 0.03$ & $0.65 \pm 0.02$ & $0.27 \pm 0.01$ & $0.08 \pm 0.01$ \\
$7.28 \pm 0.04$ & $0.64 \pm 0.02$ & $0.24 \pm 0.01$ & $0.12 \pm 0.01$ \\
$8.54 \pm 0.05$ & $0.65 \pm 0.02$ & $0.23 \pm 0.01$ & $0.11 \pm 0.01$ \\
$10.19 \pm 0.07$ & $0.55 \pm 0.01$ & $0.29 \pm 0.01$ & $0.17 \pm 0.01$ \\
$12.57 \pm 0.09$ & $0.49 \pm 0.01$ & $0.24 \pm 0.01$ & $0.27 \pm 0.01$ \\
$15.72 \pm 0.13$ & $0.48 \pm 0.01$ & $0.23 \pm 0.01$ & $0.29 \pm 0.01$ \\
$19.93 \pm 0.18$ & $0.42 \pm 0.01$ & $0.21 \pm 0.01$ & $0.37 \pm 0.01$ \\
$25.09 \pm 0.26$ & $0.35 \pm 0.01$ & $0.19 \pm 0.01$ & $0.47 \pm 0.02$ \\
$31.45 \pm 0.37$ & $0.30 \pm 0.01$ & $0.15 \pm 0.00$ & $0.55 \pm 0.02$ \\
$38.90 \pm 0.50$ & $0.28 \pm 0.01$ & $0.12 \pm 0.00$ & $0.60 \pm 0.02$ \\
$47.35 \pm 0.68$ & $0.23 \pm 0.01$ & $0.15 \pm 0.01$ & $0.62 \pm 0.02$ \\
$57.73 \pm 0.92$ & $0.19 \pm 0.01$ & $0.12 \pm 0.00$ & $0.68 \pm 0.02$ \\
$71.45 \pm 1.28$ & $0.21 \pm 0.01$ & $0.06 \pm 0.00$ & $0.73 \pm 0.02$ \\
$89.71 \pm 1.83$ & $0.20 \pm 0.01$ & $0.08 \pm 0.00$ & $0.73 \pm 0.02$ \\
\hline
\hline
\end{tabular}
\end{table}

\end{document}